\documentstyle[twocolumn,aps,psfig]{revtex}       
\newcommand{ \be }{\begin{equation}}
\newcommand{ \ee }{\end{equation}}
\newcommand{ \bea }{\begin{eqnarray}}
\newcommand{ \eea }{\end{eqnarray}}
\newcommand{ \la }{\langle}
\newcommand{ \ra }{\rangle}

\newcommand{ \bp }{{\bf p}}

\newcommand{ \bE }{{\bf E}}
\newcommand{ \bB }{{\bf B}}

\newcommand{\f}{\frac}

\newcommand{\ds}{\displaystyle}

\title{
\begin{flushright}
\vspace*{-1.8cm}
{\rm STAR Note SN0420 } 
\end{flushright}
\vspace{1.cm}
Discussing the possibility of observation of  
parity violation in heavy ion collisions}

\author{ 
Sergei A. Voloshin }                                                  

\address{ 
\vspace{0.2cm}
Department of Physics and Astronomy, Wayne State University,
Detroit, MI 48201 }
 
\begin{document}          
\maketitle
\begin{abstract}                                                              
It was recently argued that in heavy ion collision the parity could be 
broken.  
This Note addresses the question of possibility of the experimental
detection of the effect. 
We discuss how parity violating effects would modify 
the final particle distributions and how one could construct variables
sensitive to the effect, and which measurement would be the (most)
conclusive.
Discussing different observables we also discuss the question
if the ``signals'' can be faked by ``conventional'' effects (such as
anisotropic flow, etc.) and make estimates of the signals.
\end{abstract} 

\section{Introduction}

Kharzeev, Pisarski and Tytgat\cite{khar98} argue that during
the evolution of the hot (QGP) fireball created in heavy ion collision
meta-stable parity odd bubbles can be created. 
Such bubbles would have a non-zero expectation value of  
$\la \bB \cdot \bE \ra \neq 0$,
where $\bB$ and $\bE$ are the chromo-magnetic and chromo-electric
fields.
The expectation value 
$\la \bB \cdot \bE \ra$ is not sign definite and 
would take  positive and  negative values with equal probabilities.
Originally~\cite{khar98} it was proposed to look for the effect 
by detecting the non-statistical fluctuations in the variable
\be
J=\sum_{\pi^+, \pi^-} \f{\ds (\vec{p}_{\pi^+} \times \vec{p}_{\pi^-})_z}
{\ds p_{\pi^+} p_{\pi^-}}.
\ee 
Later, Gyulassy~\cite{gyul99} proposed to use for this purpose 
the so-called twist tensor:
\be
T_{ij}=\sum_{\pi^+, \pi^-}
(\vec{p}_{\pi^+} \times \vec{p}_{\pi^-})_i
(\vec{p}_{\pi^+} - \vec{p}_{\pi^-})_j .
\ee 
Other observables as well as relations between them were also discussed
in~\cite{khar99,thomas99}. 
The purpose of the current Note is not
to discuss and compare all different P- and/or CP-odd variables
(though we do discuss some of them), but instead concentrate 
on the general approaches to the question of experimental detection of
the hypothetical bubbles with parallel electric and magnetic field.
This problem clearly belongs to what now is usually called
Event-by-Event (EbyE) physics. 
The parity violating effects modify the particle distributions on the
EbyE basis and we try to apply EbyE techniques to detect the signal.
We also show that sometimes the effect of parity violation can be confused with
other effects (having nothing to do with parity violation) such as
anisotropic flow, and caution should be used analyzing different signals.

In our discussion we adopt the idea of Chikanian and 
Sandweiss~\cite{chik99}, who for simplicity proposed to 
simulate the effect of parity odd bubbles by bubbles with 
parallel (real) magnetic and electric fields randomly oriented in space.
Note that the real effect caused by color fields 
is {\sl not} necessarily opposite for positive and negative pions as it is
for the real electric and magnetic fields. Thus it is very important whenever
possible to measure the effect separately for each particle species
including baryons and anti-baryons. 
The observables discussed in this Note provide such a possibility.
 
The Note is organized as following.
The discussion of the effect of parity odd bubbles on particle
momentum distributions we split into two parts.
The effects related to transverse field component and due to 
the longitudinal component are discussed  separately.
Then based on the picture we get, we discuss how the effect can be  
observed experimentally. There exist two classes of possible
observables, being sensitive only to one of the fields or to the both
of them. We discuss both classes.
Finally we make simple statistical estimates of the signal (and background).

In our discussion we often assume that the parity odd bubble
is located at midrapidity, and we consider the effect of particle
distribution modification separately in the forward and backward
hemispheres.
In principle the bubble can be produced anywhere in rapidity, 
and the corresponding splitting of the entire rapidity space into two parts 
can be done at any rapidity point.

\section{ Effect of the transverse field components}

We start with the case of non-zero transverse component of the 
electric and magnetic fields. We choose the coordinate system such that
the magnetic field points in the $y$ direction. 
The electric field would point either in the same 
or in the opposite direction.
The effect of the fields on the particle distribution is the following. 
First, the magnetic field ``rotates'' the distribution about the $y$ axis.
Fig.~1a shows qualitatively such a rotation for positively charged particles.
Next, the electric field ``shifts'' the entire distribution along the
$y$ axis either in the positive or negative direction based on
the orientation of the field and charge of the particle (Fig.~1b). 

\begin{figure}
\centerline{\psfig{figure=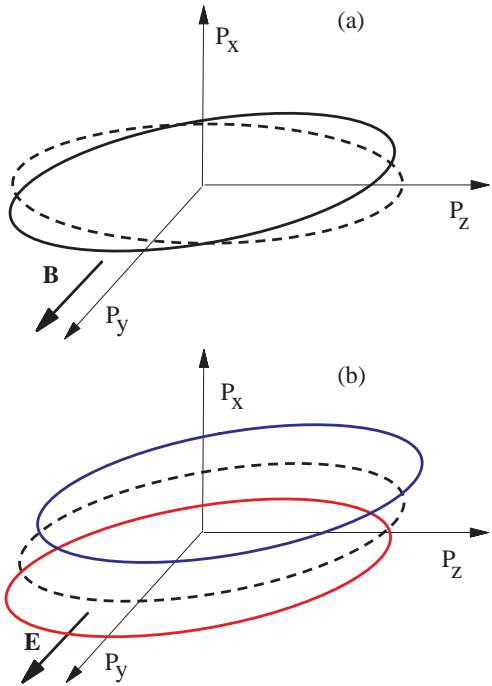,height=9.cm}}
\vspace*{5mm}
  \caption[]{(a) The rotation of the (positive) particle distribution 
due to the magnetic field. (b) The shifts of the distributions of
  positive and negative particles in the opposite directions due to
  electric field.
    }
\label{fig1}
\end{figure}                                     

How these changes in the distribution can be detected? 
The ``cleanest'' (and the most robust) observable for the effect 
would be the one which is sensitive to both fields. 
One of the simplest observable of this kind is the so-called
$V$ variable.
It is also important that this variable can be constructed 
using only one kind of particles (e.g. positive pions, protons,
anti-nucleons, etc.).
It uses the average transverse momenta of particles with positive 
and negative  rapidities (or pseudorapidities), 
$\la \bp_t \ra_{\eta>\eta_c}=(1/N_{\eta>\eta_c})\sum_{\eta>\eta_c}\bp_t$ and   
$\la \bp_t \ra_{\eta<\eta_c}=(1/N_{\eta<\eta_c})\sum_{\eta<\eta_c}\bp_t$,
where the sums run over all particles in the rapidity interval.
$N_{\eta>\eta_c}$ and $N_{\eta<\eta_c}$ are the corresponding multiplicities.
The result of the rotation of the distribution due to magnetic field 
on $\la \bp_t \ra$ is opposite in the forward and backward
hemispheres. Then the quantity 
$(\la \bp_t \ra_{\eta>\eta_c}-\la \bp_t \ra_{\eta<\eta_c})$ would be a good
measure of the strength of the magnetic field (how it is constructed
this quantity on average has nonzero ``x'' component, positive in Fig. 1.)
(If it would be real magnetic field it could be better
to weight each particle with its longitudinal momentum. We do not
discuss possible weights at this moment). 

The effect of the electric field is on the contrary
similar in both hemispheres. To ``feel'' the electric field we use 
the quantity 
$(\la \bp_t \ra_{\eta>\eta_c}+\la \bp_t \ra_{\eta<\eta_c})$ (oriented 
along the ``y'' axis in our
example if one consider positive particles).
Finally we construct the variable
\be
V = \{(\la \bp_t \ra_{\eta>\eta_c}-\la \bp_t \ra_{\eta<\eta_c}) \times 
(\la \bp_t \ra_{\eta>\eta_c}+\la \bp_t \ra_{\eta<\eta_c})\}_z
\ee
The value of $V$ is directly proportional to $\la \bB \cdot \bE \ra$ and thus
directly measures the effect. $V$ depends on both electric and
magnetic fields and thus is quadratic in the field
strengths. Due to this the effect  may be small in magnitude.
We leave the numeric estimates for the last section of the Note. 
As it was already mentioned the electric field can be either parallel
or anti-parallel to the magnetic field. It means that $V$ would have
both positive and negative values. The non-zero effect would manifest
itself by non-statistical fluctuations in $V$. It could be measured,
for example, by the sub-event method (see section on estimates
and~\cite{ebe99} for description of the method).

If the strength of the signal permits the best would be to correlate
the magnitude of $(\la \bp_t \ra_{\eta>\eta_c}+\la \bp_t
\ra_{\eta<\eta_c})$ 
to the perpendicular to it component of 
$(\la \bp_t \ra_{\eta>\eta_c}-\la \bp_t \ra_{\eta<\eta_c})$  
in order to prove that $\bB$ and $\bE$ fields are correlated, or check
that the electric and magnetic fields are indeed aligned, that is to
check if  $(\la \bp_t \ra_{\eta>\eta_c}+\la \bp_t
\ra_{\eta<\eta_c})$ is perpendicular to 
$(\la \bp_t \ra_{\eta>\eta_c}-\la \bp_t \ra_{\eta<\eta_c})$.

Let us now discuss the possibility to observe 
the {\sl first order effects}, namely the effects due to only magnetic or 
only electric field. 
We  start with {\sl magnetic} field. 
As can be seen directly from Fig.~1 the effect of the magnetic field 
(rotation about the $y$ axis and predominant particle emission in
one of the transverse directions for particles in the forward
hemisphere and in the opposite direction in the backward hemisphere) is
indistinguishable from the effect of directed flow 
(which can be small but not negligible even
for very central collisions). 
One can argue that the parity violation effect should be different
for positive and negative pions, but the same could be true for
directed flow.
Taking into account that the effect of P/CP-odd bubbles expected 
to be rather small (some estimates are given below) it would 
be extremely difficult to disentangle it from the effect of 
``conventional'' directed flow. 
Even if the effect is large one wold have to prove
that the observed effect is due to the parity violation and not to
anomalously large directed flow.

At this point one can ask why the effects are so similar, while
directed flow obviously does not violate parity. 
The answer to this question is that the directed flow can ``rotate'' 
the distribution only in the reaction plane. Any rotation in any other
plane would constitute the parity violation. 
Unfortunately, in reality we do not know the real reaction plane
orientation, and the particle azimuthal distribution itself 
is used to determine the plane. 
Then it is not at all clear what is the cause for the observed anisotropy    
in the azimuthal particle distribution.
The variable $V$ discussed above (as any other variable
sensitive to both fields, e.g. the twist tensor) correlates the
effects due to magnetic and electric fields and thus is not confused
by the anisotropic flow.  

The effect of the {\sl electric} field (shifts of the positive and negative
pion distributions in opposite directions) in principle should be
also possible  to observe, but once more one has to prove that 
is not due to Coulomb interactions, and/or resonance decays, etc..

\section{ Longitudinal field components}

Now we move on to the discussion of the effect of the longitudinal 
components of the electric and magnetic fields. 
The electric field ``shifts'' positive particles along the $z$ axis
(read, rapidity) while shifting negative particles in the opposite direction.
The magnetic field would ``rotate'' the particle distribution 
about the $z$ axis.
In principle the magnitude of the ``shift'' due to the electric field 
could be correlated with the change in particle distribution 
due to magnetic field (the correlation similar to the one discussed 
in the previous section), 
but as it is shown below the effect of magnetic field itself
would be an unambiguous signal of parity violation. 
Thus we concentrate in this section on the effects 
sensitive to only electric or magnetic field.

The {\sl electric} field effect (relative shift of the rapidity distribution
of positive and negative particles) from our point of view
can be confused with the effects due to Coulomb 
interactions and/or resonance decays, unless
the electric field effect happen to be extremely strong.
The hope here would be to observe strong EbyE fluctuations in the
shift, but once more one would have to calculate the possible
fluctuations in Coulomb fields. 
Much ``cleaner'' signal could be the one based on the effect of 
the {\sl magnetic} field, which presumably ``rotates'' the initial 
distribution 
about the $z$ axis in opposite directions for positive and negative particles.
If the initial distribution is azimuthally symmetric such a
rotation  obviously do not produce any noticeable effect and is not
detectable (as was already noticed in~\cite{chik99}).
But in the real collisions the distribution is not 
expected to be azimuthally symmetric due to directed and/or elliptic
flow! Then the magnetic field effect becomes observable.  

The direction of the rotation of the distribution is different 
for positive and negative particles
as shown in Fig. 2.
\begin{figure}
\centerline{\psfig{figure=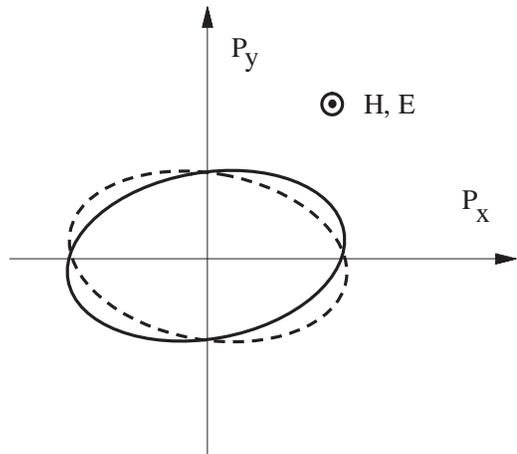,height=6.cm}}
  \caption[]{The rotation of the distribution due to longitudinal
  component of the magnetic field. 
Positive and negative particles exhibit opposite effect.
    }
\label{fig2}
\end{figure}                       
Such rotations would lead to the difference in the reaction planes
reconstructed separately using positive or negative 
particles\footnote{The procedure of the reaction plane
reconstruction now is quite well established~\cite{posk_volo98}.}.
One should have in mind that the final observable effect is a product of two, 
the anisotropic flow and the parity violation (magnetic field), 
and can be small.
Expressed as the mean sine of the azimuthal 
angle difference between
positive and negative particles in a given event
the effect is
\be
\la \sin(\phi_{\pi^+}-\phi_{\pi^-}) \ra \approx 2 v_n \la \Delta
\phi_H \ra,
\ee
where $v_n$ ($n=1,2$) is the anisotropic flow parameter (n-th Fourier
coefficient in the particle azimuthal angle distribution with respect
to the reaction plane; for definition see, for
example,~\cite{posk_volo98})
and $\la  \Delta \phi_H \ra $ is the mean (over all particles in a
given event) rotation angle due to the magnetic field.
$ \la \Delta \phi_H \ra $ can be positive or negative depending on the
orientation of the field and thus one has to study the non-statistical
fluctuations in this quantity, $\sigma_{\sin(\Delta \phi),non-stat}$. 

In the analysis, especially if one studies elliptic flow,
it could be more convenient to use the reconstructed
reaction planes, not the azimuthal angles of the individual particles.
Then for a weak signal one gets:
\bea
& & \la \sin(\Psi_{RP,\pi^+}-\Psi_{RP,\pi^-}) \ra
\nonumber
\\
& &\approx
\sqrt{N_{\pi^+} N_{\pi^-}} 
\la \sin(\phi_{\pi^+}-\phi_{\pi^-}) \ra . 
\eea

Such kind of analysis was done by the NA49
Collaboration~\cite{volo99,sikler99}
for Pb+Pb collisions at CERN SPS energies. 
In that analysis the non-statistical fluctuations in the azimuthal
angle between positive and negative pions have been measured. 
The results are presented as an upper limit
on
$\sigma_{\sin(\Delta \phi),non-stat} < 10^{-3}$, the variance 
of the angle difference. 
According to the discussion above, one has to divide this quantity
by the flow signal (in that case $v_1$) typically 
of a few percent in order to get the limit on the rotational angle due
to the bubble magnetic field.

\section{ Numeric estimates}

The impulse that acts on the particle crossing the bubble
is estimated~\cite{imp} to be about 30~MeV. It is similar for both 
electric and magnetic fields. Not all particles in the collision 
cross the bubble boundaries. The fraction would obviously 
depends on the bubble volume. In our estimates we will use
that the mean impulse due to either field is $\Delta p \approx
\alpha \cdot 30$~MeV. Then $\alpha $ would be the fraction of all 
 particles (in the acceptance) suffered a collision with the bubble boundary.
In the STAR acceptance for central Au+Au collision 
we expect about 2000 charged particles.
In our analysis one often has to subdivide this number into two 
parts (e.q. forward and backward hemispheres),
which gives about 1000 particles in each part. 
We also use an estimate (comes from RQMD) 
for $\la p_x^2\ra \approx (350~MeV)^2$.
Then the ``signal to background'' ratio in a quantity like
$\la p_x \ra_{\eta > \eta_c}$ would be of the order of 
$(\alpha \cdot 30/\sqrt{3})/(350/\sqrt{1000}) \approx 1.5 \alpha$, where we
divided the impulse by  a factor of $\sqrt{3}$, taking into
account that the direction of the corresponding field is not fixed.

All quantities discussed as a signal of parity violation are not sign
definite and one has look for non-statistical fluctuations in such
quantities. The subevent method is probably one of the best for this
purpose. This technique involves the subdivision of all particles in a
given event into two groups\footnote{It can be done in many ways, 
each of them has its own advantages and disadvantages, for  discussion
see~\cite{ebe99}.} with subsequent correlation of the signals in each
of the groups (called subevents). The number of particles in a
subevent is about half of that of the event, and signal to
background ratio would drop to $S/B \approx \alpha$.
Having in mind that one needs to correlate the subevents we get in the
correlation function $S/B \approx \alpha^2$. The last step in this
direction would be to take into account the event statistics. Then
$S/B \approx \alpha^2 \sqrt{N_{events}}$.

The above estimates are relevant mostly for a variables such as $V$
variable.
For the correlation of the reaction planes the relevant quantity would
be 
\be
\la \theta_H \ra 
\approx \Delta p / \sqrt{3} /\la p_t\ra 
\approx  \alpha \cdot 30/\sqrt{3} /(\sqrt{2} \cdot 350)
\approx 0.05 \alpha.
\ee
The anisotropic flow parameters are (at SPS) of the order of
$v_n\approx 0.02-0.06$. 
Then for $\sigma_{\sin(\Delta \phi),non-stat}$ one would expect 
values about
\be
\sigma_{\sin(\Delta \phi),non-stat} \approx \alpha 
\cdot (1-3)\cdot10^{-3}.
\ee 
Remind you that the NA49 preliminary limit on this quantity is
$<10^{-3}$.

\section{Conclusion}
 
Parity violation in strong interactions is a question of a fundamental
value. The experimental detection of the effect is a challenge and a
perfect example of a problem  of Event-by-Event physics.
The search is expected to be difficult, but as discussed in this
Note as well as in~\cite{chik99,thomas99} it is not hopeless in a sense
that results valuable for theory can be obtained. 
  
We should probably also mention here a ``homework'' for theorists.
In the case the effect would be experimentally observed one would have
to prove that it is not due to large fluctuations of the real electric and
magnetic fields. Theoretical estimates of such fluctuations in the
volume of the fireball created in heavy ion collision are highly desirable.

\section*{Acknowledgements}

Discussions with all members of the STAR parity group, and in particular
with A.~Chikanian, J.~Sandweiss, J.~Thomas, as well as with D.~Kharzeev 
are greatly acknowledged. I also thank I. Sakrejda for useful comments.

This work was supported by the Director, Office of Energy Research,
Office of High Energy and Nuclear Physics, Division of Nuclear Physics
of the U.S. Department of Energy under Contracts 
DE-FG02-92ER40713.

Many of  references can be found at 
http://www.star.bnl.gov/STAR/html/parity\_l/index.html.

\end{document}